\documentclass[preprint,3p,times]{elsarticle}
\usepackage{amssymb}
\usepackage{lipsum}

\usepackage{hyperref}
\usepackage[T1]{fontenc}

\newcommand{\pt}{p_\mathrm{T}}
\newcommand{\elab}{E_{lab} = 6.7,~8,~11,~\mathrm{and~25~A~GeV}}
\newcommand{\particles}{\mathrm{\pi^+,~K^+,~p,~K_s^0,~\Lambda, ~and~\Xi^- }}
\newcommand{\mesons}{\mathrm{\pi^+,~K^+,~and~K_s^0}}
\newcommand{\baryons}{\mathrm{p,~\Lambda, ~and~\Xi^- }}

\begin{document}

\begin{frontmatter}

\title{Collision energy dependence of elliptic flow of identified hadrons in heavy-ion collisions using the PHSD model}

\author[1]{B. Towseef\corref{cor1}}
\ead{bhattawseef82@gmail.com}
\cortext[cor1]{Corresponding Author}
				
\author[1]{M. Farooq}
				
\author[2]{V. Bairathi\corref{cor2}}
\ead{vipul.bairathi@gmail.com}
\cortext[cor2]{Corresponding Author}
				
\author[1]{B. Waseem}
				
\author[2]{S. Kabana}
				
\author[1]{S. Ahmad}
				
\affiliation[1]{organization={University of Kashmir},
				addressline={Srinagar},
				postcode={190006},
				country={India}}
\affiliation[2]{organization={Instituto de Alta Investigación, Universidad de Tarapacá},
				addressline={Casilla 7D},
				city={Arica},
				postcode={1000000},
				country={Chile}}
				
\begin{abstract}
We report the first predictions of elliptic flow ($v_2$) of identified hadrons at mid-rapidity ($|y| <$ 1.0) in Au+Au collisions at $\elab$ using the Parton Hadron String Dynamics (PHSD) model. The transverse momentum ($\pt$) dependence of identified hadron $v_2$ in different centrality intervals (0-10\%, 10-40\%, and 40-80\%) are shown. A clear centrality dependence of $v_2(\pt)$ is observed for particles at $E_{lab} = $ 25 and 11 A GeV, while the centrality dependence becomes weaker for particles at $E_{lab} = $ 8 and 6.7 A GeV. Within the PHSD model, the number of constituent quark (NCQ) scaling of $v_{2}$ approximately follows in Au+Au collisions at all beam energies. A collision energy ($\sqrt{s_{NN}}$) dependence of $\pi$, $K$, and $p$ $v_2$ is studied in comparison with available published experimental data in the beam energy range of 6-25 A GeV. These predictions will help in interpreting the data from the forthcoming Compressed Baryonic Matter (CBM) experiment at the Facility for Antiproton and Ion Research (FAIR) and Multi-Purpose Detector (MPD) at the Nuclotron-based Ion Collider facility (NICA).
\end{abstract}

\begin{keyword}
Identified hadrons \sep Elliptic flow \sep Heavy-ion collisions \sep Quark-Gluon Plasma
\PACS 25.75.Ld
\end{keyword}

\end{frontmatter}

\section{Introduction}
\label{sec:Intro}
In relativistic heavy-ion collisions, a unique form of matter known as Quark Gluon Plasma (QGP), characterized by de-confined quarks and gluons, is formed~\cite{r1}. The study of QGP provides valuable insights into the properties of matter in extreme temperature and density conditions, resembling those of the early universe. QGP has been extensively investigated through experiments at the Relativistic Heavy Ion Collider (RHIC) and the Large Hadron Collider (LHC), focusing on high temperatures ($T$) and low net-baryon densities ($\mu_B$)~\cite{r2,r3,r4,r5,r6}. Current research in heavy-ion collision experiments is centered on the phase transition from hadronic matter to QGP and the Quantum Chromodynamics (QCD) phase diagram at finite net-baryon density. An experimental approach to understand the phase structure of nuclear matter in a moderate baryon density region involves varying collision energies and studying observables as a function of collision centrality, transverse momentum ($\pt$), and rapidity ($y$). 

One of the most widely studied observable in heavy-ion collisions is the azimuthal anisotropy of the produced particles. It provides valuable insight into the expansion dynamics and properties of the matter produced in heavy-ion collisions~\cite{r7,r8}. Azimuthal anisotropies result from irregularities in the initial overlapping region and fluctuations in nucleon positions during non-central heavy-ion collisions, leading to asymmetries in particle production relative to the reaction plane~\cite{r9,r10}. The distribution of produced particles with respect to the angle of the reaction plane ($\Psi_{RP}$) is commonly described using a Fourier series~\cite{r11,r12}:
\begin{equation}
E\frac{d^3N}{dp^3} = \frac{d^2N}{2\pi p_{T} dp_{T} dy} \left[ 1 + 2\sum_{n=1}^{\infty}v_{n}\cos(n(\phi - \Psi_{RP})) \right],
\label{eq:1}
\end{equation}
where $\phi$ represents the azimuthal angle of the particle. The coefficients $v_{n}$, known as flow coefficients, are used to quantify the level of azimuthal anisotropy in particle production. Of these coefficients, $v_{2}$, the second-order coefficient, is specifically referred to as the elliptic flow.

At high energies, when the colliding nuclei transit time is shorter than the typical particle production time, the elliptic flow is mainly influenced by the collective expansion of the initial partonic density distribution~\cite{r13}. This conclusion is based on the observed number of constituent quark (NCQ) scaling of identified hadron $v_{2}$ in the intermediate $\pt$ range of 2 to 5 GeV/c~\cite{r14,r15}. The quark number scaling at top RHIC energy has been studied within the PHSD model~\cite{r15b}. Conversely, at low energies, the observed elliptic flow is significantly affected by the initial state baryon stopping and the nuclear mean-field effects~\cite{r16,r17,r18}. In this case, a deviation from the NCQ scaling of identified hadrons $v_{2}$ could indicate the absence of the partonic phase. Various calculations of hydrodynamic and microscopic transport models have highlighted the significance of elliptic flow at low beam energies~\cite{r19,r20,r21,r22,r23}. Thus, it is essential to study the beam energy dependence of $v_{2}$ to differentiate between hadronic and partonic phases and comprehend the properties of the medium generated in heavy-ion collisions.

Several beam energy scan programs are actively investigating the phase transitions within QCD. Notable among these are the Beam Energy Scan (BES) program at the RHIC, covering energy ranges from $\sqrt{s_{NN}} =$ 3-39 GeV~\cite{r24,r25,r26,r27}. Additionally, the NA61/SHINE experiment at the Super Proton Synchrotron (SPS) explores energies between $\sqrt{s_{NN}} =$ 5.1-17.3 GeV, while the HADES experiment at the SIS-18 facility of GSI focuses on $\sqrt{s_{NN}} =$ 2.4-2.55 GeV~\cite{r17,r18,r28}. These programs aim to uncover the characteristics of the phase transition from partonic to hadronic matter and identify the critical point of QCD. In future, CBM experiment at the FAIR and MPD at the NICA facility, will explore the QCD phase diagram in the region of high baryon density. The CBM experiment will be conducted at a center of mass energy of $\sqrt{s_{NN}} =$ 2.7-4.9 GeV, while the MPD will be conducted at $\sqrt{s_{NN}} =$ 4-11 GeV~\cite{r29,r30}. 

This work investigates the collective behavior of the QGP medium produced in heavy-ion collisions at FAIR and NICA energies using the PHSD transport model~\cite{r31,r32}. The identified hadron $v_2$ of $\pi$, $K$, $K^{0}_{s}$, $p$, $\Lambda$, and $\Xi$ at mid-rapidity ($|y| <$ 1.0) in Au+Au collisions at $\elab$ will be reported. The centrality and transverse momentum dependence of identified hadron $v_2$ will be discussed. We have also examined the constituent quark number scaling of $v_{2}(\pt)$. Furthermore, we have investigated the $v_{2}$ integrated over $\pt$ and $y$ as a function of the center of mass energy and compared our results with published data from various experiments. These findings are the first predictions of identified hadrons $v_{2}$ at beam energies $\elab$ using the PHSD model, and they are particularly relevant for the future CBM experiment at FAIR and MPD experiment at NICA.
 
\section{PHSD model}
\label{sec:model}
The PHSD model is a microscopic covariant dynamical approach used to study strongly interacting systems, both in and out of equilibrium~\cite{r31,r32}. It is based on the Dynamical Quasi-Particle Model (DQPM) and formulated using Kadanoff-Baym equations for Green's functions in phase-space representation~\cite{r33,r34,r35,r36}. The PHSD model describes the full evolution of a relativistic heavy-ion collision from the initial hard scatterings and string formation through the dynamical de-confinement phase transition to the strongly interacting QGP as well as hadronization and the subsequent interactions in the expanding hadronic phase. It includes both effective partonic and hadronic degrees of freedom, where the field quanta are described by dressed propagators with complex self-energies. Once the proper complex self-energies of the degrees of freedom are known, the off-shell transport equations for quarks and hadrons fully control the time evolution of the system. In the PHSD model, heavy quarks are produced by hadron string decay above a critical energy density of 0.5 GeV/$fm^3$, and quark fusion in hadronization results in the production of hadrons. The PHSD depicts the coexistence of the quark-hadron mixture at energy densities close to the critical energy density. The PHSD approach is equivalent to the Hadron-String-Dynamics (HSD) model in the hadronic phase, for energy densities below the critical energy density. The details of the PHSD model and their comparison with experimental observables in heavy-ion collisions from the SPS to RHIC energies are available in Refs.~\cite{r37,r38,r39,r40,r41}.

\section{Analysis details}
\label{sec:analysis}
\begin{figure}[htbp]
\centering 
\includegraphics[width=0.5\textwidth]{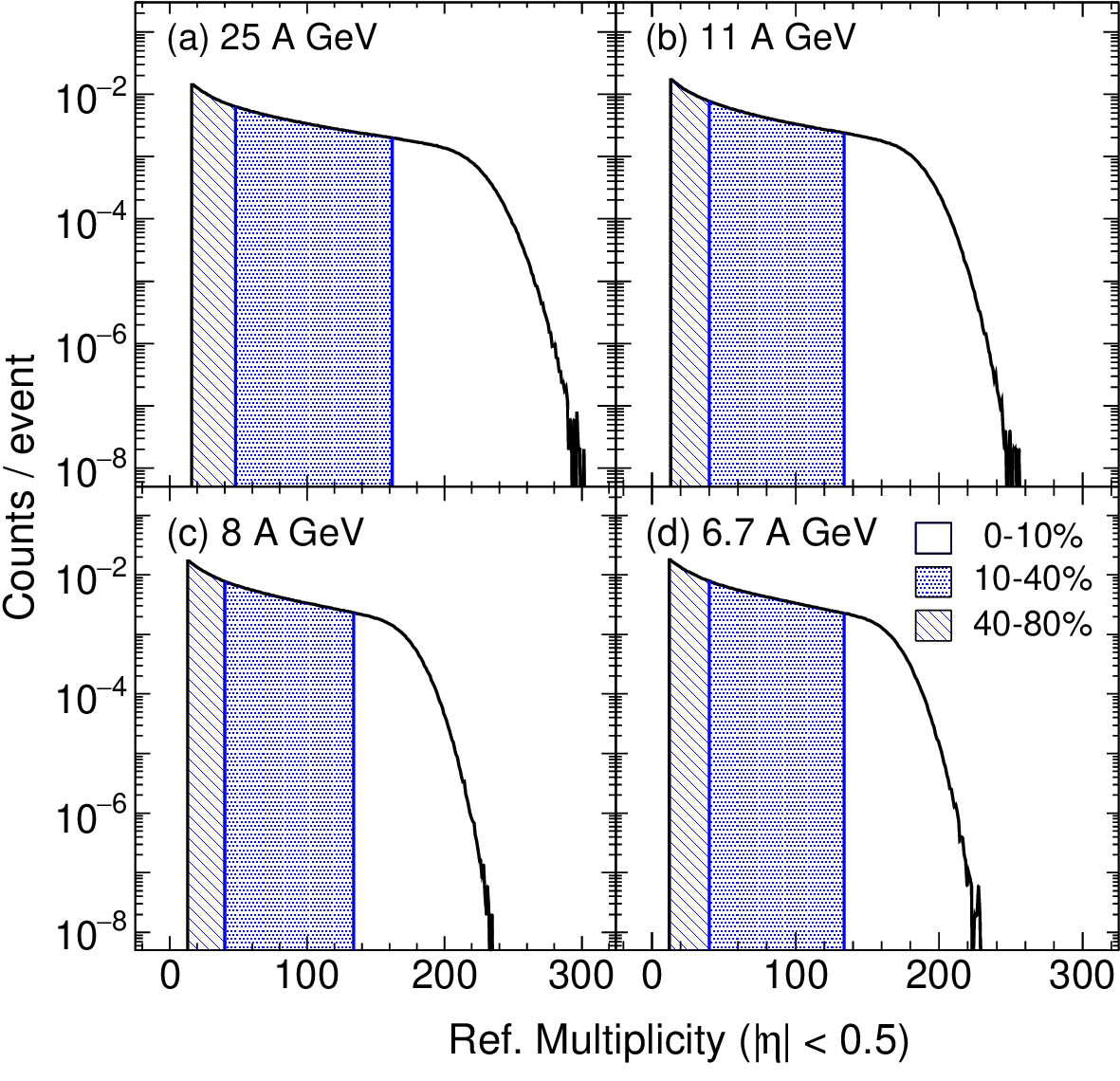}
\caption{(Color online) Reference multiplicity distribution in Au+Au collisions for $\elab$ from the PHSD model. The bands represent 0-10\%, 10-40\%, and 40-80\% centrality intervals for each energy.}
\label{fig:RefMult}
\end{figure}

The results presented in this paper are based on the version 4.1 of the PHSD model. A data set of 50 million minimum bias Au+Au collision events is generated in the impact parameter (b) range between 0 to 15 fm for beam energies $\elab$. The $v_2$ calculations in this study are performed at different centrality intervals that cover central to peripheral collisions. The centrality of an event is determined based on the reference multiplicity calculated within a pseudorapidity range of $|\eta| < 0.5$ for all energies from the PHSD model. The reference multiplicity is divided into three centrality classes: 0-10\% (central), 10-40\% (mid-central), and 40-80\% (peripheral), as shown in Fig.~\ref{fig:RefMult}.

The PHSD model has a fixed orientation for the impact parameter along the x-axis and the beam direction along the z-axis. Consequently, the reaction plane is aligned in the x-z direction by default at $0$ degree. Keeping the experimental approach in mind, the second-order event plane angle $\psi_{2}$, an estimate of the reaction plane angle, is reconstructed event-by-event using charged hadrons in the pseudorapidity range $|\eta| < 1.0$ and transverse momentum range $0.2 < \pt < 2.0$~GeV/c. The elliptic flow $v_{2}$ of identified hadrons is then calculated with respect to $\psi_2$ using the standard event plane method described in Ref.~\cite{r12} as follows:
\begin{equation}
v_2 = \langle \cos[2(\phi - \psi_2)] \rangle.
\label{eq:2}
\end{equation} 
The finite number of charged hadrons in each event limits the estimation of the true reaction plane angle. As a result, the $v_{2}$ is corrected using the event plane resolution ($R_{2}$), which is determined by the $\eta$-sub event plane method. In this method, the two independent sub-events are defined using charged hardons in the negative ($-1 < \eta < -0.05$) and positive ($0.05 < \eta < 1$) pseudorapidity regions. The event plane resolution is then calculated as follows:
 \begin{equation}
R_2 = \sqrt{\langle \cos[2(\psi_2^a - \psi_2^b)] \rangle }.
\label{eq:3}
\end{equation}
Here, the $\langle..\rangle$ denote an average over all particles and events. $\psi_{2}^{a}$ and $\psi_{2}^{b}$ are the event plane angles in the negative and positive pseudorapidity sub-events, respectively. This method reduces non-flow effects using an $\eta$ gap 0.1 between the two independent sub-events. 

\begin{figure}[htbp]
\centering 
\includegraphics[width=0.4\textwidth]{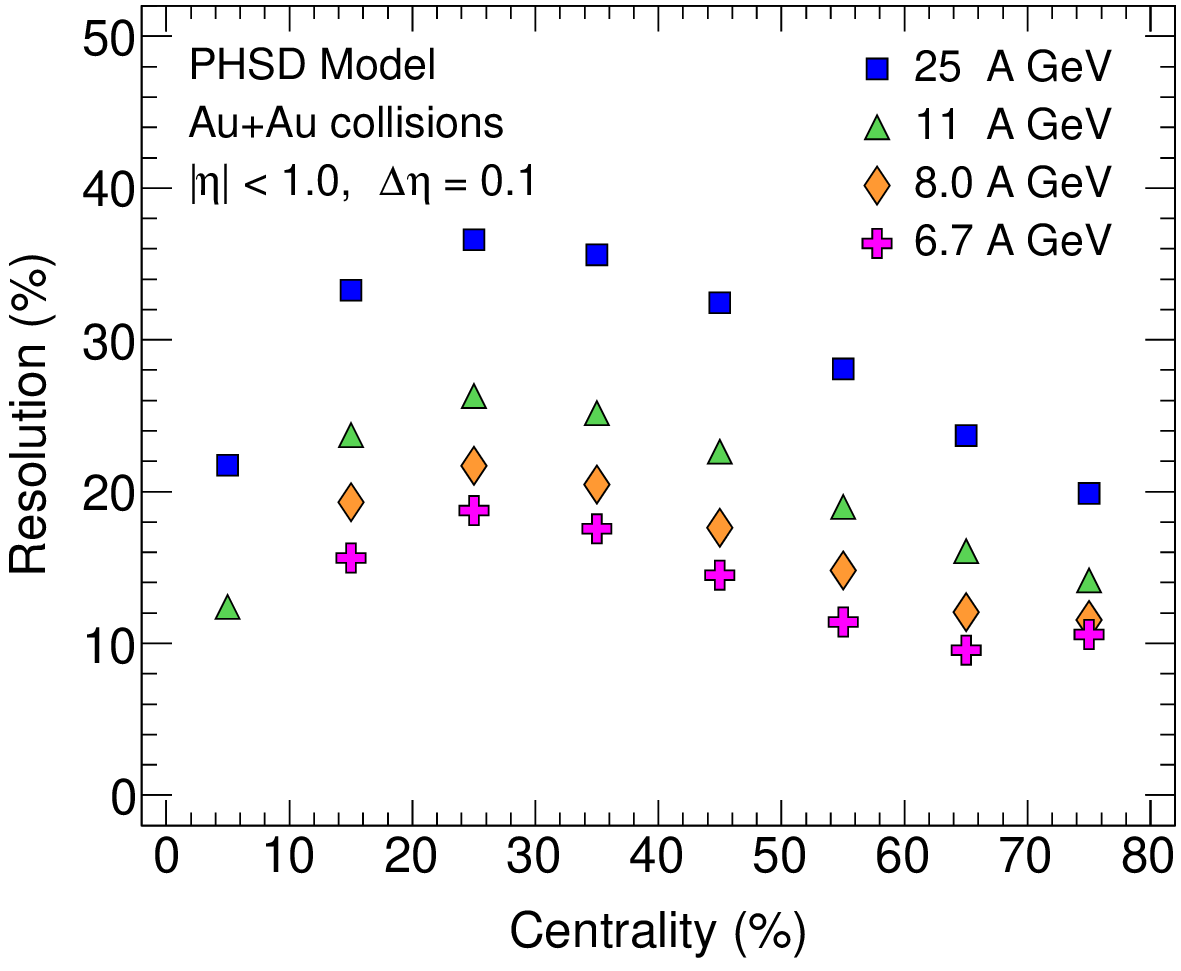}
\caption{(Color online) Event plane angle resolution as a function of centrality using the $\eta$-sub event plane method in Au+Au collisions at $\elab$ from the PHSD model.}
\label{fig:EpRes}
\end{figure}

Figure~\ref{fig:EpRes} shows the event plane angle resolution as a function of collision centrality in Au+Au collisions at $\elab$ from the PHSD model. The resolution is highest for 25 A GeV and decreases with decreasing beam energy due to the lower particle multiplicities and flow magnitudes. For the lowest two beam energies, 6.7 and 8 GeV, the calculations of $v_2$ are not possible for most central (0-10\%) collisions due to negative resolution values using this method. To minimize the auto-correlation effects, the elliptic flow for particles in the positive pseudorapidity window is calculated with respect to the $\psi_{2}$ calculated in the negative pseudorapidity window, and vice versa. The elliptic flow computed using Eq.~\ref{eq:2} is then divided by the event plane resolution ($R_{2}$) to get the final flow coefficient.

\section{Results and Discussion}
\label{sec:results}
\subsection{Differential elliptic flow $v_2(\pt)$}
\label{sec:v2pt}
\begin{figure}[htbp]
\centering 
\includegraphics[width=0.6\textwidth]{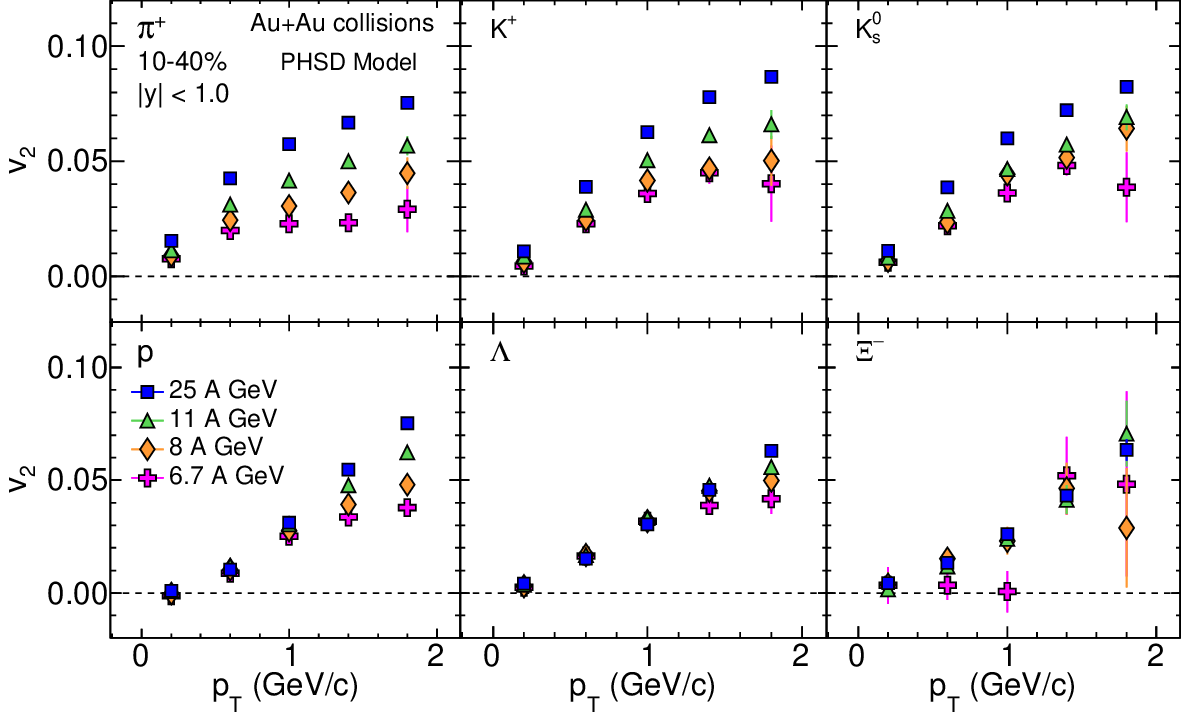}
\caption{(Color online) Differential $v_2(\pt)$ for identified hadrons ($\particles$) in 10-40\% central Au+Au collisions at mid-rapidity for beam energies $\elab$ from the PHSD model. The statistical uncertainties are indicated by the error bars.}
\label{fig:v2ptmb}
\end{figure}

We present differential $v_2$ as a function of $\pt$ for identified hadrons ($\particles$) at mid-rapidity ($|y| <$ 1.0) in 10-40\% central Au+Au collisions at $\elab$ from the PHSD model as shown in Fig.~\ref{fig:v2ptmb}. We observe that the $v_{2}$ of all particles exhibits a similar dependence on $\pt$ in all beam energies. The $v_2(\pt)$ of mesons ($\mesons$) increases with increasing beam energy, showing a clear dependence on beam energy, whereas baryons ($\baryons$) show relatively less beam energy dependence than the mesons. This could be due to the increasing contribution of transported quarks from the initial state baryons in the mid-rapidity region. Moreover, it should be noted that as the beam energy decreases, the baryon chemical potential of the system at chemical freeze-out increases, as mentioned in Ref.~\cite{r42,r43}. This increase could also be a contributing factor to the higher $v_{2}$ of baryons compared to mesons at lower beam energies.

\subsection{Centrality dependence of $v_2(\pt)$}
\begin{figure}[htbp]
\centering 
\includegraphics[width=0.6\textwidth]{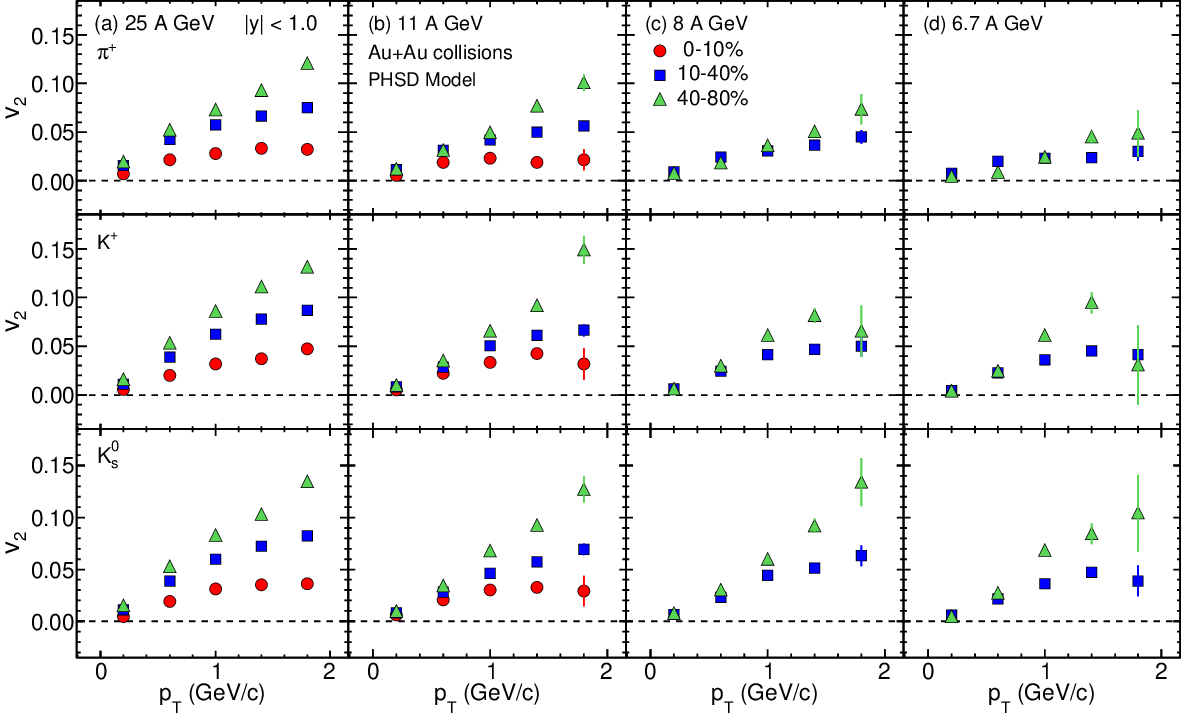}
\caption{(Color online) Elliptic flow $v_2(\pt)$ of mesons ($\mesons$) in 0-10\%, 10-40\%, and 40-80\% Au+Au collisions at $\elab$ from the PHSD model. The error bars are statistical uncertainties.}
\label{fig:v2pt_cent1}
\end{figure}

\begin{figure}[htbp]
\centering 
\includegraphics[width=0.6\textwidth]{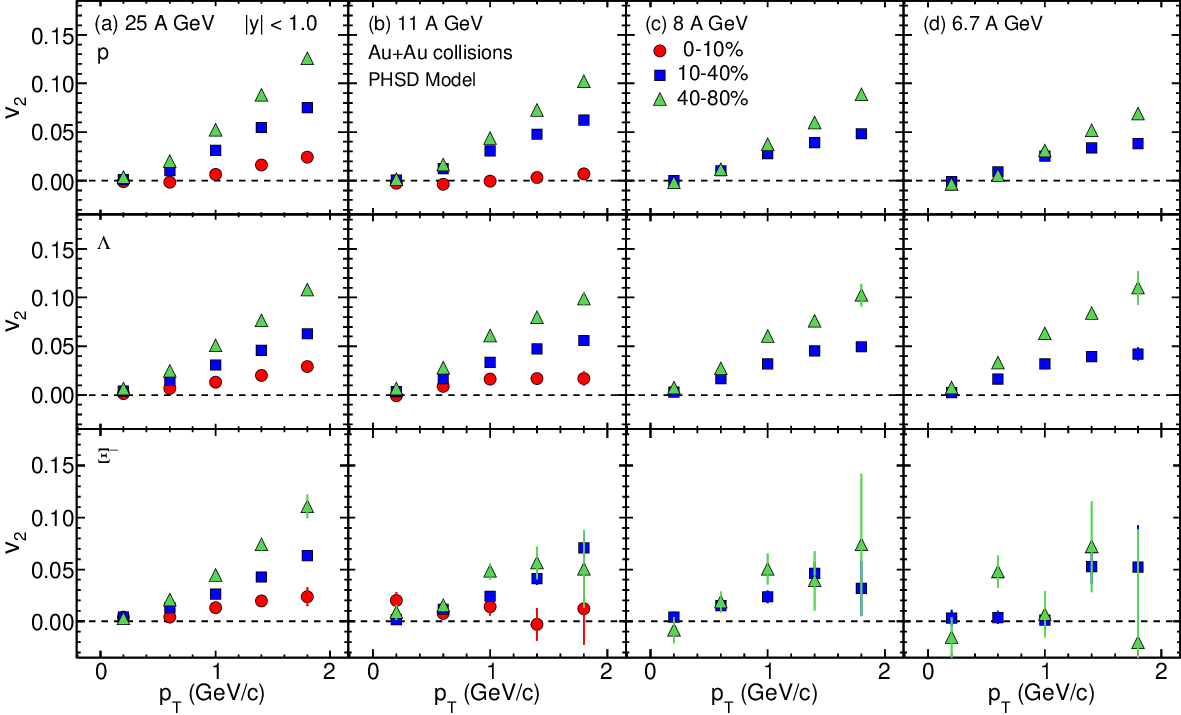}
\caption{(Color online) Elliptic flow $v_2(\pt)$ of baryons ($\baryons$) in 0-10\%, 10-40\%, and 40-80\% Au+Au collisions at $\elab$ from the PHSD model. The error bars are statistical uncertainties.}
\label{fig:v2pt_cent2}
\end{figure}

The centrality dependence of $v_2$ in heavy-ion collisions gives insights into the collision dynamics, collision geometry, and particle production mechanism~\cite{r44,r45,r46}. Therefore, we present centrality dependence of $v_{2}(\pt)$ for mesons ($\mesons$) and baryons ($\baryons$) for $\elab$ from the PHSD model in Fig.~\ref{fig:v2pt_cent1} and \ref{fig:v2pt_cent2}, respectively. We observe a clear centrality dependence of $v_2$ for mesons and baryons at $E_{lab}$ = 25 and 11 A GeV, whereas a weak centrality dependence is observed for $E_{lab}$ = 8 and 6.7 A GeV. The $v_2$ values for the most central collision (0-10\%) are lower than those for the peripheral collisions (40-80\%). The dependence of $v_2$ on the collision centrality reflects the interplay between initial collision geometry and particle production in final state from central to peripheral collisions~\cite{r45,r46}.

\subsection{Number of constituent quark scaling}
\label{sec:NCQ}
\begin{figure}[htbp]
\centering 
\includegraphics[width=0.8\textwidth]{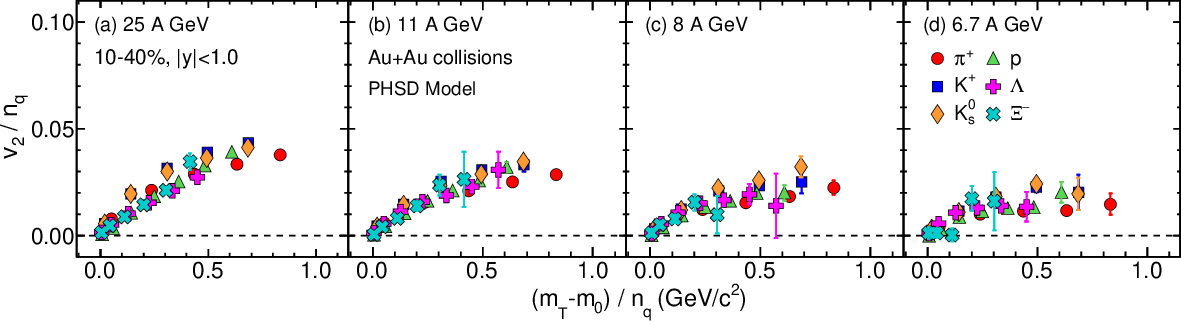}
\caption{(Color online) The NCQ scaled $v_2/n_q$ as a function of scaled transverse kinetic energy ($m_T - m_0$)/$n_q$ for identified hadrons ($\particles$) in 10-40\% central Au+Au collisions for beam energies $\elab$ from the PHSD model. The error bars are statistical uncertainties.}
\label{fig:NCQ1}
\end{figure}

The observation of constituent quark scaling is considered critical evidence for the formation of QGP in high-energy heavy-ion collisions~\cite{r47}. This scaling arises from the collective behavior of partons produced during the early stages of the collisions~\cite{r14,r15,r48,r49,r50}. It is based on the recombination and coalescence of partons as described in various transport models~\cite{r51,r52,r53}. In order to explore the NCQ scaling, we report $v_2$ scaled by number of constituent quarks $n_q$ (3 for baryons and 2 for mesons) for various beam energies using the PHSD model. Figure~\ref{fig:NCQ1} shows $v_2/n_q$ versus transverse kinetic energy ($m_T - m_0$)/$n_q$ for identified hadrons ($\particles$) in 10-40\% central Au+Au collisions at $\elab$. The NCQ scaling seems to follow within statistical uncertainties at these low beam energies with moderate $\mu_B$ in the PHSD model. Notably, previous measurements of identified hadron $v_2$ by the STAR experiment at RHIC at $\sqrt{s_{NN}}$ = 4.5 GeV also demonstrate a similar NCQ scaling~\cite{r26}. This study using the PHSD model further reveals that NCQ scaling of $v_2$ is following even at the lowest studied beam energy, indicating partonic collectivity at $E_{lab}$ = 6.7 A GeV ($\approx\sqrt{s_{NN}}$ = 4.0 GeV).

\subsection{Beam energy dependence of $v_2$}
\label{sec:energy_dependence}
The study of the beam energy dependence of $v_2$ is crucial for understanding the transition from hadronic to partonic phase in heavy-ion collisions. Studies suggested that the positive $v_2$ with NCQ scaling is a result of strong partonic expansion in the early stages of high energy heavy-ion collisions. In contrast, the negative $v_2$ and absence of scaling at lower energies may be attributed to baryon stopping and shadowing of the spectators. Measurements performed by the STAR experiment at RHIC in Au+Au collisions at $\sqrt{s_{NN}} = 3.0$ GeV indicate that baryonic interactions play a dominant role in the collision dynamics. Furthermore, a comparison of the $v_2$ results with JAM and UrQMD transport models, incorporating baryonic mean-field, also supports the conclusion that baryon interactions are the dominant degrees of freedom at $\sqrt{s_{NN}} = 3.0$ GeV~\cite{r27}.
\begin{figure}[htbp]
\centering 
\includegraphics[width=0.45\textwidth]{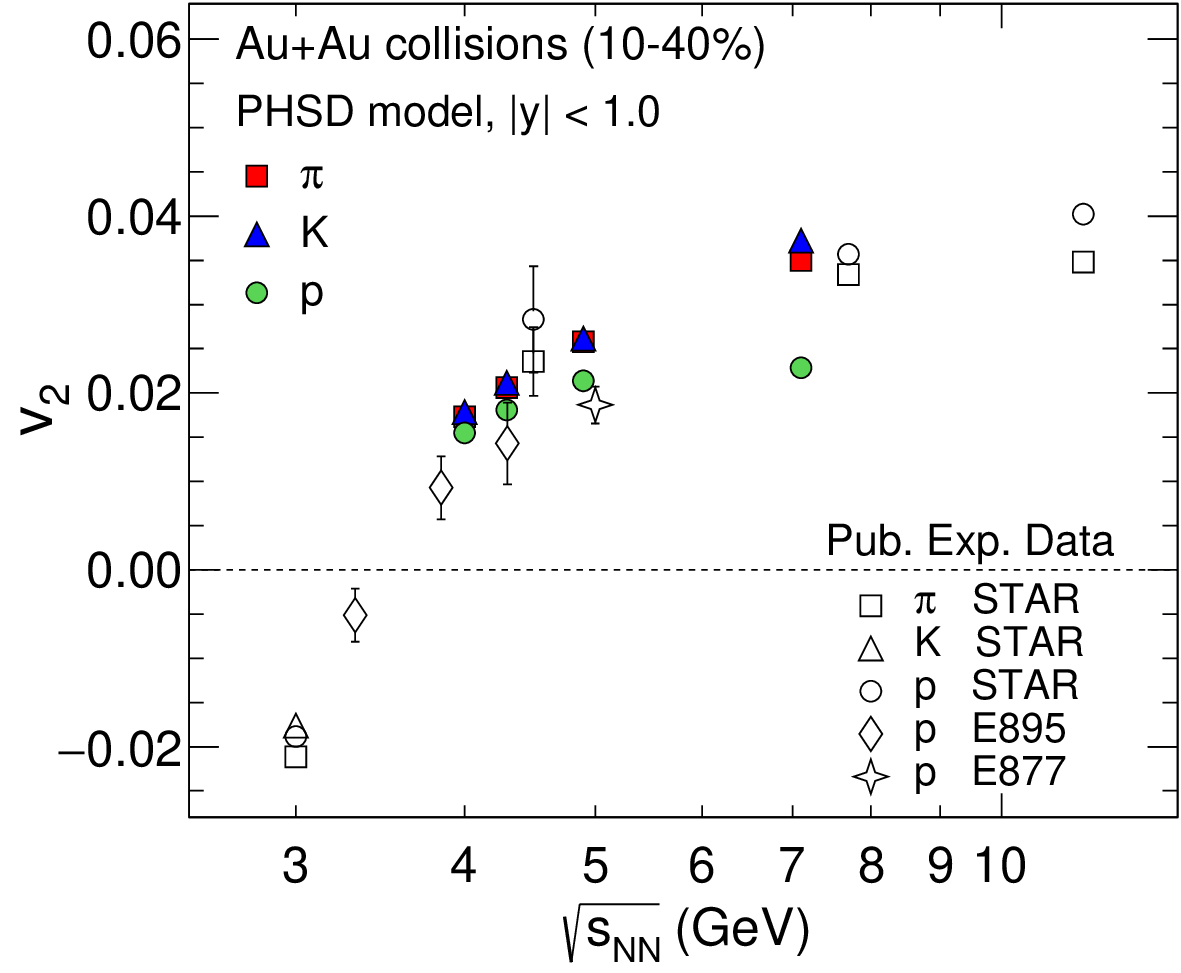}
\caption{(Color online) $v_2$ as a function of $\sqrt{s_{NN}}$ for $\pi$ (combined $\pi^{\pm}$), $K$ (combined $K^{\pm}$), and $p$ in 10-40\% central Au+Au collisions from the PHSD model. The $v_2$ results from the E877, E895 and STAR experiments in Au+Au collisions are also shown by open markers\cite{r16,r24,r26,r27}. The error bars are statistical uncertainties.}
\label{fig:v2E}
\end{figure}

In this section, we investigate the beam energy dependence of the elliptic flow of identified hadrons. Figure~\ref{fig:v2E} shows the $v_2$ integrated over $p_T$ (0.2 $< \pt <$ 2.0 GeV/c for $\pi, K$ and 0.4 $< \pt <$ 2.0 GeV/c for $p$) and rapidity $|y| < 1.0$ as a function of $\sqrt{s_{NN}}$ for $\pi$, $K$, and $p$ in 10-40\% central Au+Au collisions from the PHSD model. The calculated $v_2$ values are compared with the existing published $v_2$ results measured in Au+Au collisions by various experiments ~\cite{r16,r24,r26,r27}. It is noted that $v_2$ increases with an increase in beam energy. The $v_2$ values for pions and protons are in line with the published experimental results, except at the highest beam energy $E_{lab} =$ 25 A GeV for protons. This increase of $v_2$ with beam energy in the PHSD model is mainly due to the increasing influence of the partonic interactions via repulsive partonic mean-field potential and by parton scattering while at low energies the building of $v_2$ is mainly due to the hadronic interactions via an attractive scalar hadron-hadron potential and a repulsive momentum dependent vector potential as well as by hadronic scatterings~\cite{r54}. 

\section{Summary}
\label{sec:summary}
In summary, the elliptic flow of identified hadrons ($\particles$) at mid-rapidity in Au+Au collisions from the PHSD model using the $\eta$-sub event plane method have been reported. The study focuses on the beam energy dependence of the identified hadron $v_2$ in the energy range $\elab$. A clear beam energy dependence of $v_2(\pt)$ for mesons ($\mesons$) is observed. However, baryons ($\baryons$) show weak energy dependence in the 6-25 A GeV energy range. This could be due to more number of initial state baryon transported to the mid-rapidity region with decreasing beam energies. Furthermore, the $v_2$ of mesons and baryons shows a clear centrality dependence at $E_{lab}$ = 25 and 11 A GeV, while a weak centrality dependence is observed for $E_{lab}$ = 8 and 6.7 A GeV. Additionally, the study explores the NCQ scaling behavior of $v_2$ as a function of transverse kinetic energy. The NCQ scaling of $v_2$ seems to approximately follow within statistical uncertainties at these beam energies in the PHSD model, indicating the formation of a hydrodynamic medium with dominant partonic degrees of freedom. The collision energy ($\sqrt{s_{NN}}$) dependence of $v_2$ is also studied for $\pi$, $K$ and $p$. The $v_2$ of pion and proton from the PHSD model are found to be comparable to the results from the E895, E877, and STAR experiment at RHIC. These predictions of identified hadrons $v_{2}$ from the PHSD model are particularly relevant for the future CBM experiments at FAIR and MPD experiments at NICA in the beam energies range from 6-25 A GeV.
 
\section*{Acknowledgements}
\label{acknowledgement}
S. Kabana and V. Bairathi acknowledge the financial support received by ANID PIA/APOYO AFB220004. This research was supported in part by the cluster computing resource provided by the IT Division at the GSI Helmholtzzentrum für Schwerionenforschung, Darmstadt, Germany. The authors acknowledge helpful advice from the PHSD group members E. L. Bratkovskaya, V. Voronyuk, W. Cassing, P. Moreau, O. 
E. Soloveva, and L. Oliva.

\end{document}